# Multi-pooling 3D Convolutional Neural Network for fMRI Classification of Visual Brain States

Zhen Zhang
*School of Informatics*
*Kochi University of Technology*
Kochi, Japan
248002m@gs.kochi-tech.ac.jp

Masaki Takeda
*Research Center for Brain Communication*
*Kochi University of Technology*
Kochi, Japan
takeda.masaki@kochi-tech.ac.jp

Makoto Iwata
*School of Informatics*
*Kochi University of Technology*
Kochi, Japan
iwata.makoto@kochi-tech.ac.jp

*Abstract*—Neural decoding of visual object classification via functional magnetic resonance imaging (fMRI) data is challenging and is vital to understand underlying brain mechanisms. This paper proposed a multi-pooling 3D convolutional neural network (MP3DCNN) to improve fMRI classification accuracy. MP3DCNN is mainly composed of a three-layer 3DCNN, where the first and second layers of 3D convolutions each have a branch of pooling connection. The results showed that this model can improve the classification accuracy for categorical (face vs. object), face sub-categorical (male face vs. female face), and object sub-categorical (natural object vs. artificial object) classifications from 1.684% to 14.918% over the previous study [1] in decoding brain mechanisms.

## I. Introduction

fMRI is a non-invasive and reliable technique that measures the small changes in blood flow caused by brain activities. fMRI data have relatively high spatial resolution and availability, thus providing a solution to reveal neural activities of brain regions under visual stimuli. Recent years witnessed the applications of convolutional neural networks (CNNs) in brain decoding, such as [1][2][3][4].

The previous study [1] performed categorical (face vs. object), face sub-categorical (male face vs. female face), and object sub-categorical (natural object vs. artificial object) classifications via a classic three-layer 3DCNN, revealing that the human visual system recognizes objects following the principle of going from categories into sub-categories.

However, the classification model used in [1] does not significantly present a high accuracy even with 9-fold fMRI data averaging, especially for sub-categorical classification tasks. Therefore, a novel multi-pooling 3D convolutional neural network (MP3DCNN) is proposed in this paper, which is expected to reach a higher accuracy than the previous one [1] and play a valuable role in decoding brain mechanisms.

## II. Materials

### A. fMRI data collection and preprocessing

In our study, the fMRI dataset is provided by [1], where the subject clicked the button corresponding to a random visual stimulus (an image of a male face, female face, natural object, or artificial object) within 0.5s. During this period, a Siemens 3T MRI scanner recorded the subject's brain states as T1-weighted 3D fMRI volumes. Through SPM12 [5], the fMRI volumes were realigned, co-registered, normalized to the standard Montreal Neurological Institute (MNI) template, and resampled to 2-mm isotropic voxels.

### B. Available fMRI data

There are 17306 fMRI volumes of size (79 × 95 × 79) from 50 subjects available, including 4453, 4399, 4214, and 4240 volumes corresponding to the visual stimuli of male face, female face, natural object, and artificial object, respectively. To suppress the background noise and the irrelevant neural activities, the fMRI dataset of each subject was multi-fold averaged as an option to improve the data quality (for example 9-fold fMRI data averaging).

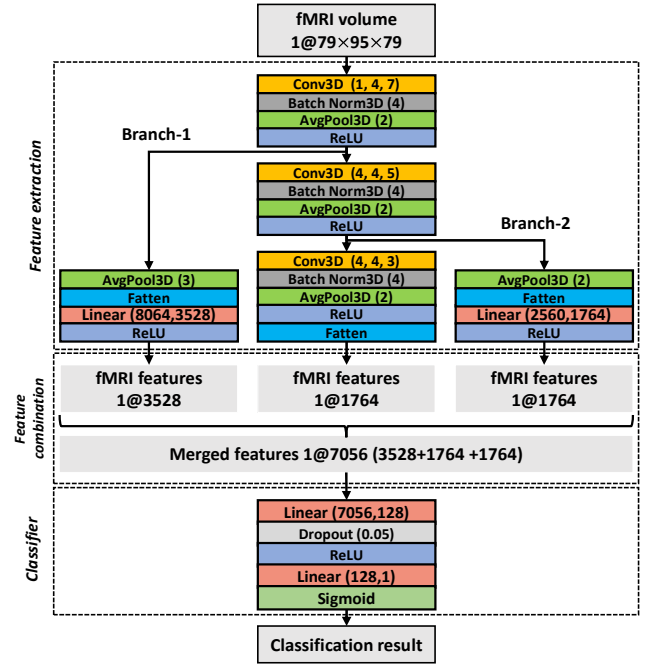

Fig.1 Multi-pooling 3D convolutional neural network (MP3DCNN).

## III. Methods

The proposed MP3DCNN includes feature extraction, feature combination, and classifier, as shown in Fig.1.

### A. Feature extraction

The feature extraction has a mainchain and two branches. Its mainchain is a three-layer 3D CNN, and each convolution combines with a batch normalization, an average 3D pooling layer, and a rectified linear unit (ReLU). The first and second 3D convolutions each have a branch connection, where an average 3D pooling layer and a linear layer are used to generalize further and connect the extracted features. The relevant hyperparameters are as follows:

**3D Convolution.** An 3D convolution is a 3D tensor with the hyperparameter of ($c_{in}$, $c_{out}$, $k_{conv}$), where $c_{in}$ depends on the number of the input feature maps (default 1 to the input fMRI volume), $c_{out}$ determines the number of the convolutional filter, and $k_{conv}$ is the kernel size of the convolutional calculation. Batch normalization is applied on $c$ dimension, where $c$ equals to the $c_{out}$ of the former-layer convolution. The hyperparameters of the three convolutions in the feature extraction are (1,4,7), (4,4,5) and (4,4,3), respectively.

**Average 3D pooling.** Suppose the $i_{th}$ 3D convolution outputs a 3D tensor with the size of $c_i \times h_i \times w_i \times d_i$, where $c_i$ is the number of the feature maps, $h_i$, $w_i$, and $d_i$ indicate the height, width and depth of each feature map. An average 3D pooling with a hyperparameter of $k_{pool}$ can perform average pooling on each feature map using a kernel size of $k_{pool}$, which is set to 2 in the mainchain. In the branch-1 and -2, $k_{pool}$ is set to 3 and 2.

**Linear layer.** Linear layer with the hyperparameters of $(f_{in}, f_{out})$ can convert 1D tensors of arbitrary size. The output of the mainchain is flattened into a 1D tensor with the size of $1 \times 1764$. In the branch-1 and 2, the hyperparameters of the linear layer are $(8064, 3528)$ and $(2569, 1764)$, which aim to output the same number and the double number of features as the mainchain.

*B. Feature combination and Classifier*

In the feature extraction unit, the branch-1, branch-2, and mainchain output the first, the second, and the third-level feature representations of fMRI data, respectively. Through feature combination, the model can provide a sophisticated decision using multi-level features in fMRI classification. Moreover, the branch connections added extra paths for the first and the second convolutions in the backpropagation, aiming to improve the model's learning efficiency.

The classifier is composed of two linear layers with the hyperparameters of $(5292, 128)$ and $(128, 1)$. The dropout can prevent overfitting in the training processing, and the probability is set to 0.05. Sigmoid activation is used for binary classification tasks.

IV. EXPERIMENTS AND RESULTS

The proposed model was implemented by Pytorch [6] on a Linux platform with an NVIDIA GeForce RTX 3060 GPU. We trained and tested the proposed and previous models using unaveraged and 9-fold averaged fMRI volumes. To evaluate the model performance independently, the fMRI volumes from 5 of 50 subjects were detached as the test dataset. According to the study [1], we equalized the test dataset and retained 1468 (734 faces vs. 734 objects), 772 (386 male faces vs. 386 female faces), and 718 (359 natural objects vs. 359 artificial objects) of fMRI volumes for categorical, face sub-categorical, and object sub-categorical classifications. The fMRI volumes of the 45 leftover subjects were used as the training dataset.

In the model training process, we performed 9-fold cross-validation with 25 training iterations, where the batch size was 64, the learning rate was 0.00001, and the loss function was binary cross entropy (BCE) [7]. Within the 25 iterations, the model parameters with the highest accuracy score on each validation dataset were used for model testing. Finally, we used a majority voting scheme to ensemble the nine results in determining the classification accuracy.

Table I presents the comparison of the voting accuracy between the proposed and the previous models [1] for the categorical, face sub-categorical, and object sub-categorical classifications. We can see that the proposed model improved the voting accuracy by 14.918%, 3.368%, and 2.646% when using unaveraged fMRI volumes, and 4.088%, 1.684%, and 13.649 % when using 9-fold averaged fMRI volumes over the original model in the three classifications.

TABLE I
COMPARISON BETWEEN THE PROPOSED AND PREVIOUS MODELS [1]

|  | Voting accuracy | |
| --- | --- | --- |
|  | Proposed | Previous |
| Face vs. Object (no avg) | **71.458%** | 56.54% |
| Face vs. Object (9-fold avg) | **89.578%** | 85.49% |
| Male face vs. Female face (no avg) | **52.72%** | 49.352% |
| Male face vs. Female face (9-fold avg) | **55.311%** | 53.627% |
| Natural object vs. Artificial object (no avg) | **51.532%** | 48.886% |
| Natural object vs. Artificial object (9-fold avg) | **61.978%** | 48.329% |

*__Bold:__ Better score in comparison between the proposed and previous models

V. CONCLUSION AND DISCUSSION

A novel fMRI classification model called MP3DCNN is proposed in our study. Its mainchain is a three-layer 3DCNN, where the first and the second layers of convolution each have a branch connection. In the mainchain and branches, the extracted fMRI featured are multiple pooled. The results show that the proposed model reached higher classification accuracies than the previous study [1] for the categorical, face sub-categorical, and object sub-categorical classifications.

The main reasons are speculated as follows: 1) we considered that the multiple 3D average pooling layers can against the local feature redundancy in the feature extraction process and pass the global information to the classifier as much as possible; 2) through the branch connections, the model decision can depend on the merged features of the three 3D convolutions, thereby improving the model's robustness.

In future research, we look forward to using grid-search [8] to optimize the model hyperparameters and exploring the visual explanations based on the reached classification results.

ACKNOWLEDGMENTS

The authors would like to express their sincere gratitude for the invaluable help from Mr. Kosuke Miyoshi, and Dr. Shinichi Yoshida.

REFERENCES

[1] N. Watanabe et al., "Multi-modal deep neural decoding of visual object representation in humans," *bioRxiv.*, 2022.

[2] R. J. Meszlényi, K. Buza, and Z. Vidnyánszky, "Resting state fMRI functional connectivity-based classification using a convolutional neural network architecture," *Frontiers Neuroinform.*, vol. 11, p. 61, Oct. 2017.

[3] M. N. I. Qureshi, J. Oh, and B. Lee, ''3D-CNN based discrimination of schizophrenia using resting-state fMRI,'' *Artif. Intell. Med.*, vol. 98, pp. 10–17, Jul. 2019.

[4] Y. Zhao et al., ''Automatic recognition of fMRI-derived functional networks using 3-D convolutional neural networks,'' *IEEE Trans. Biomed. Eng.*, vol. 65, no. 9, pp. 1975–1984, Sep. 2018.

[5] J. Ashburner et al., "SPM12 Manual," in *Proc. Wellcome Trust Centre Neuroimag.*, London, U.K., 2014, p. 2464.

[6] A. Paszke et al., "PyTorch: An imperative style, high-performance deep learning library," in *Proc. Adv. Neural Inform. Process. Syst.*, 2019.

[7] A. U. Ruby, P. Theerthagiri, I. J. Jacob, and Y. Vamsidhar, "Binary cross entropy with deep learning technique for image classification," *Int. J. Adv. Trends Comput. Sci. Eng.*, vol. 9, no. 4, pp. 5393–5397, Aug. 2020.

[8] S. M. LaValle, M. S. Branicky, and S. R. Lindemann, "On the relationship between classical grid search and probabilistic roadmaps," *Int. J. Robot. Res.*, vol. 23, no. 7, pp. 673–692, Jul.–Aug. 2004.